\documentstyle[aps,12pt]{revtex}
\begin{document}

\title{Interdimensional degeneracies \\
for a quantum $N$-body system in $D$ dimensions}

\author{Xiao-Yan Gu \thanks{Electronic address:
guxy@mail.ihep.ac.cn} and Zhong-Qi Ma \thanks{Electronic address:
mazq@sun.ihep.ac.cn}}

\address{CCAST (World Laboratory), P.O.Box 8730, Beijing 100080, China \\
and Institute of High Energy Physics, Beijing 100039, China}

\author{Jian-Qiang Sun \thanks{Electronic address:
sunjq@mail.ihep.ac.cn}}

\address{Institute of High Energy Physics, Beijing 100039, China}

\maketitle

\date{}

\vspace{5mm}

\begin{abstract}

Complete spectrum of exact interdimensional degeneracies for a
quantum $N$-body system in $D$-dimensions is presented by the
method of generalized spherical harmonic polynomials. In an
$N$-body system all the states with angular momentum $[\mu+n]$ in
$(D-2n)$ dimensions are degenerate where $[\mu]$ and $D$ are given
and $n$ is an arbitrary integer if the representation $[\mu+n]$
exists for the SO($D-2n$) group and $D-2n\geq N$. There is an
exceptional interdimensional degeneracy for an $N$-body system
between the state with zero angular momentum in $D=N-1$ dimensions
and the state with zero angular momentum in $D=N+1$ dimensions.

\end{abstract}

\vspace{10mm} For a quantum few-body system in $D$ dimensions, one
of the characteristic features is the presence of exact
interdimensional degeneracies. Perhaps first noticed by Van Vleck
\cite{van}, an isomorphism exists between angular momentum $l$ and
dimension $D$ such that each unit increment in $l$ is equivalent
to two-unit increment in $D$ for any central force problem in $D$
dimensions. For a two-body system (e.g. one-electron atom) states
related by the dimensional link $D,~l \leftrightarrow
(D-2),~(l+1)$ are exactly degenerate \cite{her2,her3}. For
three-body system (e.g. two-electron atom) Herrick and Stillinger
found exact interdimensional degeneracies between the states
$^{1,3}P^e$ and $^{1, 3}D^o$ in $D=3$ and the states $^{3, 1}S^e$
and $^{3, 1}P^o$ in $D=5$, respectively \cite{her3}. For a
four-body system (e.g. three-electron atom) Herrick \cite{her2}
found an exceptional interdimensional degeneracy that the triply
excited $2p^{3}$ $^{4}S$ fermion state of the lithium atom is
exactly degenerate with the spinless boson $1s^{3}$ ground state
for $D=5$ ($D=5$ was misprinted as $D=3$ in Ref. \cite{her2}). In
1961 Schwartz \cite{sch} proved by the recursion relation that for
a three-body system in three-dimensional space, any angular
momentum state can be expanded in a complete set of the
independent bases whose number is finite. Recently, by the method
of the generalized Schwartz expansion \cite{dun2,dun3}, Dunn and
Watson showed some exact interdimensional degeneracies of
two-electron system in an arbitrary $D$-dimensional space
\cite{dun4,dun5}. To our knowledge, no theoretical method has yet
dealt with interdimensional degeneracies when $N>3$.

Recently, we proved the Schwartz expansion again by the method of
generalized spherical harmonic polynomials, and presented a new
development for separating completely the global rotational
degrees of freedom from the internal ones for the $N$-body
Schr\"{o}dinger equation in three-dimensional space \cite{gu1} as
well as in D dimensions \cite{gu2,gu3}. We found a complete set of
base functions for angular momentum in the system. Any wave
function with a given angular momentum can be expanded with
respect to them where the coefficients, called the generalized
radial functions, depend only upon the internal variables. The
generalized radial equations satisfied by the generalized radial
functions are derived from the Schr\"{o}dinger equation without
any approximation \cite{gu1}. The exact interdimensional
degeneracies in a three-body system \cite{gu4} were obtained
directly from the generalized radial equations. In this Letter we
study interdimensional degeneracies for an $N$-body system in
$D$-dimensional space.

For a quantum $N$-body system in an arbitrary $D$-dimensional
space, we denote the position vectors and the masses of $N$
particles by ${\bf r}_{k}$ and by $m_{k}$, $k=1,~2,\ldots,~N$,
respectively. $M=\sum_{k} m_{k}$ is the total mass. The
Schr\"{o}dinger equation for the $N$-body system with a
spherically symmetric potential $V$ is
$$- \displaystyle {1 \over 2} \displaystyle \sum_{k=1}^{N}~
\displaystyle m_{k}^{-1} \bigtriangledown^{2}_{{\bf r}_{k}} \Psi
+V \Psi =E \Psi , \eqno (1) $$

\noindent where $\bigtriangledown^{2}_{{\bf r}_{k}}$ is the
Laplace operator with respect to the position vector ${\bf
r}_{k}$. For simplicity, the natural units $\hbar=c=1$ are
employed throughout this Letter. Replace the position vectors
${\bf r}_{k}$ with the Jacobi coordinate vectors ${\bf R}_{j}$:
$$ {\bf R}_{0}=M^{-1/2}\displaystyle \sum_{k=1}^{N}~m_{k}
{\bf r}_{k},~~~~~ {\bf R}_{j}=\left(\displaystyle {m_{j+1}
M_{j}\over M_{j+1}}  \right)^{1/2} \left({\bf
r}_{j+1}-\displaystyle \sum_{k=1}^{j}~ \displaystyle {m_{k}{\bf
r}_{k}\over M_{j}}\right), $$
$$1\leq j \leq (N-1),~~~~~~M_{j}
=\displaystyle \sum_{k=1}^{j}~m_{k},~~~~~~M_{N}=M,  \eqno (2) $$

\noindent where ${\bf R}_{0}$ describes the position of the center
of mass, ${\bf R}_{1}$ describes the mass-weighted separation from
the second particle to the first particle, ${\bf R}_{2}$ describes
the mass-weighted separation from the third particle to the center
of mass of the first two particles, and so on. In the
center-of-mass frame, ${\bf R}_{0}=0$, the $N$-body
Schr\"{o}dinger equation reduces to a differential equation with
respect to $(N-1)$ Jacobi coordinate vectors ${\bf R}_{j}$:
$$\bigtriangledown^{2}\Psi^{[\mu]}_{\bf M}({\bf R}_{1},\ldots,{\bf R}_{N-1})
=-2\left\{E-V\left(\xi\right)\right\} \Psi^{[\mu]}_{\bf M}({\bf
R}_{1},\ldots,{\bf R}_{N-1}),$$
$$\bigtriangledown^{2}=\displaystyle
\sum_{j=1}^{N-1}~\bigtriangledown^{2}_{{\bf R}_{j}} ~,\eqno (3) $$

\noindent where $[\mu]$ stands for the angular momentum as
discussed later.

In a $D$-dimensional space it needs $(D-1)$ vectors to determine
the body-fixed frame. When $D\geq N$, all Jacobi coordinate
vectors are used to determine the body-fixed frame, and all
internal variables can be chosen as
$$\xi_{jk}={\bf R}_{j}\cdot {\bf R}_{k},~~~~~~
1\leq j\leq k \leq N-1. \eqno (4) $$

\noindent We call the set of internal variables (4) the first set.
The numbers of the rotational variables and the internal variables
are $(N-1)(2D-N)/2$ and $N(N-1)/2$, respectively. When $D<N$, only
$(D-1)$ Jacobi coordinate vectors are involved to determine the
body-fixed frame, and the first set of internal variables is not
complete because it could not distinguish two configurations with
different directions of, say ${\bf R}_{D}$ reflecting to the
superplane spanned by the first $(D-1)$ Jacobi coordinate vectors.
In this case we need to use the second set of internal variables:
$$\xi_{jk}={\bf R}_{j}\cdot {\bf R}_{k},~~~~~~
\zeta_{\alpha}
=\displaystyle \sum_{a_{1}\ldots a_{D}}~\epsilon_{a_{1}\ldots a_{D}}
R_{1a_{1}}\ldots R_{(D-1)a_{D-1}}R_{\alpha a_{D}}, $$
$$1\leq j \leq D-1,~~~~~~j\leq k \leq N-1,~~~~~~D\leq \alpha \leq N-1.
 \eqno (5) $$

\noindent
The numbers of the rotational variables and the internal
variables are $D(D-1)/2$ and $D(2N-D-1)/2$, respectively.

For an $N$-body system in $D$-dimensions, the angular momentum is
described by an irreducible representation of SO($D$). When $D\geq
N$ the irreducible representation is denoted by an $(N-1)$-row
Young pattern $[\mu]\equiv [\mu_{1},\mu_{2},\ldots,\mu_{N-1}]$,
$\mu_{1}\geq \mu_{2}\geq \ldots \geq \mu_{N-1}$. Due to the
traceless condition, the representation $[\mu]$ of SO($D$) exists
only if the sum of boxes of the first two columns on the left of
the Young pattern $[\mu]$ is not larger than $D$. Some selfdual
representations, antiselfdual ones, and the equivalent ones may
occur when $N\leq D\leq 2(N-1)$. They only change the explicit
forms of the base functions. The reader is suggested to refer our
previous paper for detail \cite{gu3}.

Due to the rotational symmetry, one only needs to discuss the
eigenfunctions of angular momentum with the highest weight. The
independent base function for the angular momentum $[\mu]$ with
the highest weight is $Q^{[\mu]}_{(q)}({\bf R}_{1},\ldots{\bf
R}_{N-1})$ where $(q)$ contains $(N-1)(N-2)/2$ parameters
$q_{jk}$, $1\leq k\leq j\leq N-2$, and determines a standard Young
tableau. A Young tableau is obtained by filling the digits $1$,
$2$, $\ldots$, $N-1$ arbitrarily into a given $(N-1)$-row Young pattern
$[\mu]$. A Young tableau is called standard if the digit in every
column of the tableau increases downwards and the digit in every
row does not decrease from left to right. The parameter $q_{jk}$
denotes the number of the digit "$j$" in the $k$th row of the
standard Young tableau. $q_{jk}$ should satisfy the
following constraints:
$$\begin{array}{ll}
\displaystyle \sum_{j=k}^{r}~q_{jk} \leq \displaystyle
\sum_{j=k-1}^{r-1}~q_{j(k-1)},~~~~~~
&\mu_{k+1}\leq \displaystyle \sum_{j=k}^{N-2}~q_{jk} \leq \mu_{k},\\
1\leq k \leq N-2, &k\leq r \leq N-2.
\end{array} \eqno (6) $$

\noindent The number of the independent base functions
$Q^{[\mu]}_{(q)}({\bf R}_{1},\ldots{\bf R}_{N-1})$ is equal to the
dimension $d_{[\mu]}[SU(N-1)]$ of the irreducible representation
$[\mu]$ of the SU($N-1$) group.

The explicit form of $Q^{[\mu]}_{(q)}({\bf R}_{1},\ldots{\bf
R}_{N-1})$ for the given standard Young tableau $(q)$ is very easy
to write. In the Young tableau, in correspondence to each column
with the length $t$, filled by digits $j_{1}<j_{2}<\ldots <j_{t}$,
$Q^{[\mu]}_{(q)}({\bf R}_{1},\ldots{\bf R}_{N-1})$ contains a
determinant as a factor. The $r$th row and $s$th column in the
determinant is $R_{j_{r}(2s-1)}+iR_{j_{r}(2s)}$, where ${\bf
R}_{ja}$ is the $a$th component of ${\bf R}_{j}$, if $D>2(N-1)$.
$Q^{[\mu]}_{(q)}({\bf R}_{1},\ldots{\bf R}_{N-1})$ also contains a
numerical coefficient for convenience. When $N\leq D\leq 2(N-1)$,
the explicit form of $Q^{[\mu]}_{(q)}({\bf R}_{1},\ldots{\bf
R}_{N-1})$ is a little bit changed \cite{gu3}, but it will not
affect the generalized radial equations as well as the
interdimensional degeneracies. When $D<N$, only the first $(D-1)$
Jacobi coordinate vectors are involved in the base functions
$Q^{[\mu]}_{(q)}({\bf R}_{1},\ldots{\bf R}_{D-1})$, which are the
same as those for smaller $N=D$. $Q^{[\mu]}_{(q)}({\bf
R}_{1},\ldots{\bf R}_{N-1})$ is a homogeneous polynomial of
degrees $\sum_{k} q_{jk}$ and $\sum_{j} \mu_{j}-\sum_{jk}q_{jk}$
with respect to the components of respectively the Jacobi
coordinate vectors ${\bf R}_{j}$ and ${\bf R}_{N-1}$, and
satisfies the generalized Laplace equations
$$\bigtriangledown_{{\bf R}_{j}}\cdot \bigtriangledown_{{\bf R}_{k}}
Q^{[\mu]}_{(q)}({\bf R}_{1},\ldots{\bf R}_{N-1})=0,~~~~~~ 1\leq
j\leq k \leq N-1. \eqno (7) $$

There is a one-to-one correspondence between base functions
$Q^{[\mu]}_{(q)}({\bf R}_{1},\ldots{\bf R}_{N-1})$ and
$Q^{[\mu+n]}_{(q')}({\bf R}_{1},\ldots{\bf R}_{N-1})$, where
$[\mu]\equiv [\mu_{1},\ldots,\mu_{N-1}]$, $[\mu+n]\equiv
[\mu_{1}+n,\ldots,\mu_{N-1}+n]$, and
$q_{jk}^{\prime}=q_{jk}+n\delta_{jk}$. As a matter of fact, each
standard Young tableau for $[\mu+n]$ can be obtained from a
corresponding standard Young tableau for $[\mu]$ by adhering from
its left $n$ columns with $N-1$ rows where the boxes in the $j$th
row are filled with $j$, $1\leq j \leq N-1$. From viewpoint of
group theory, two representation $[\mu]$ and $[\mu+n]$ of
SU($N-1$) are equivalent to each other and their dimensions
$d_{[\mu+n]}[SU(N-1)]=d_{[\mu]}[SU(N-1)]$.

When $D\geq N$, any wave function
$\Psi^{[\mu]}_{\bf M}({\bf R}_{1},\ldots,{\bf R}_{N-1})$
with the given angular momentum $[\mu]$ can be expanded
with respect to the complete and independent base functions
$Q^{[\mu]}_{(q)}({\bf R}_{1},\ldots,{\bf R}_{N-1})$
$$\Psi^{[\mu]}_{\bf M}({\bf R}_{1},\ldots,{\bf R}_{N-1})
=\displaystyle \sum_{(q)}~\psi^{[\mu]}_{(q)}(\xi)
Q^{[\mu]}_{(q)}({\bf R}_{1},\ldots,{\bf R}_{N-1}). \eqno (8) $$

\noindent The coefficients $\psi^{[\mu]}_{(q)}(\xi)$, called the
generalized radial functions, only depends upon the internal
variables. When $D<N$, $\psi^{[\mu]}_{(q)}(\xi)$ and
$Q^{[\mu]}_{(q)}({\bf R}_{1},\ldots,{\bf R}_{N-1})$ in Eq. (8)
have to be replaced with $\psi^{[\mu]}_{(q)}(\xi, \zeta)$ and
$Q^{[\mu]}_{(q)}({\bf R}_{1},\ldots,{\bf R}_{D-1})$, respectively.
Substituting Eq. (8) into the $N$-body Schr\"{o}dinger equation
(3), one is able to obtain the generalized radial equations. The
main calculation is to apply the Laplace operator to the wave
function $\Psi^{[\mu]}_{\bf M}({\bf R}_{1},\ldots,{\bf R}_{N-1})$.
The calculation consists of three parts. The first part is to
apply the Laplace operator to the generalized radial functions
$\psi^{[\mu]}_{(q)}(\xi)$ which can be calculated by replacement
of variables. When $D\geq N$, we have
$$\bigtriangledown^{2} \psi^{[\mu]}_{(q)}(\xi)
= \left\{\displaystyle \sum_{j=1}^{N-1}~
\left(4\xi_{jj}\partial^{2}_{\xi_{jj}}
+2 D\partial_{\xi_{jj}} \right)\right. ~~~~~~~~~~~~~~~~~~~~~~~~~~~~~~~~~~~
~~~~~~~~~~~~~~~$$
$$~~~+\displaystyle \sum_{j=1}^{N-1}\sum_{k=j+1}^{N-1}~
\left[\left(\xi_{jj}+\xi_{kk}\right)\partial^{2}_{\xi_{jk}}
+4\xi_{jk}\left(\partial_{\xi_{jj}}+\partial_{\xi_{kk}}\right)
\partial_{\xi_{jk}}\right] $$
$$\left.+2\displaystyle \sum_{j=1}^{N-1}\sum_{j\neq k=1}^{N-1}
\sum_{j\neq
t=k+1}^{N-1}~\xi_{kt}\partial_{\xi_{jk}}\partial_{\xi_{jt}}
\right\}\psi^{[\mu]}_{(q)}(\xi),~~~~~~~~~~~ \eqno (9) $$

\noindent where $\xi_{jk}=\xi_{kj}$ and $\partial_{\xi}$ denotes
$\partial/\partial \xi$ and so on. The second part is to apply the
Laplace operator to the generalized spherical harmonic polynomials
$Q^{[\mu]}_{(q)}({\bf R}_{1},\ldots,{\bf R}_{N-1})$, which is
vanishing due to Eq. (7). The third part is the mixed application:
$$2\displaystyle \sum_{j=1}^{N-1}~\left\{
\left(\partial_{\xi_{jj}}\psi^{[\mu]}_{(q)}\right)2{\bf R}_{j}
+\displaystyle \sum_{j\neq k=1}^{N-1}\left(\partial_{\xi_{jk}}
\psi^{[\mu]}_{(q)}\right){\bf R}_{k}\right\} \cdot
\bigtriangledown_{{\bf R}_{j}} Q^{[\mu]}_{(q)}.  \eqno (10) $$

\noindent
The second term is invariant under transformation $\mu_{j}
\longrightarrow \mu_{j}+n$ and $q_{jk}\longrightarrow q_{jk}+n\delta_{jk}$.
The first term is equal to
$$ Q^{[\mu]}_{(q)} \left\{4\displaystyle \sum_{j=1}^{N-2}~\left(\displaystyle
\sum_{k=1}^{j}~
q_{jk}\right)\partial_{\xi_{jj}}\psi^{[\mu]}_{(q)}(\xi)
+4\left(\displaystyle \sum_{j=1}^{N-1}~\mu_{j}- \displaystyle
\sum_{j=1}^{N-2}\sum_{k=1}^{j}~
q_{jk}\right)\partial_{\xi_{(N-1)(N-1)}}\psi^{[\mu]}_{(q)}(\xi)\right\}.
\eqno (11) $$

\noindent Under the above transformation it produces the
additional terms to the generalized radial equations
$$\displaystyle \sum_{j=1}^{N-1}~4n
\partial_{\xi_{jj}}\psi^{[\mu]}_{(q)}(\xi), \eqno (12) $$

\noindent which exactly cancel with the additional term from Eq.
(9) if $D$ is replaced with $D-2n$ at the same time.

From the above proof we come to the conclusion for the complete
spectrum of the exact interdimensional degeneracies for an
arbitrary $N$-body system with a spherically symmetric potential
that all the states in the system with the angular momentum
$[\mu+n]$ in $(D-2n)$ dimensions are degenerate where $[\mu]$
and $D$ are given and $n$ is an arbitrary integer if the
representation $[\mu+n]$ exists for the SO$(D-2n$) group and
$D-2n\geq N$, because those states are described by the wave functions
with the same number of the generalized radial functions depending
upon the same set of internal variables and satisfying the same
generalized radial equations.

Now, we turn to discuss the case of $D<N$. The base functions and
the internal variables for $D<N$ depend upon $D$ and are very
different to those for $D\geq N$ \cite{gu1,gu3} so that, generally
speaking, there is no interdimensional degeneracy when $D<N$ but
only one exception when $D=N-1$.

When $D=N-1$, $\zeta_{D}$ in the second set of internal variables
happens to be proportional to $Q^{[\mu]}_{(q)}({\bf R}_{1},\ldots
,{\bf R}_{N-1})$ with $\mu_{j}=1$, $q_{kt}=\delta_{kt}$, $1\leq j
\leq D$, and $1\leq t\leq k \leq D-1$. We denotes this
$Q^{[\mu]}_{(q)}({\bf R}_{1},\ldots ,{\bf R}_{N-1})$ as $Q_{0}$.
Note that $Q_{0}$ corresponds to a standard Young pattern of one
column with $D$ rows describing the identity representation of
SO($D$). Due to the traceless condition the Young pattern for
$Q_{0}$ is the only Young pattern with $D$ rows for SO($D$).
$\zeta_{D}^{2}$ can be expressed by the first set of internal
variables. If a wave function with zero angular momentum for an
$N$-body system in $D=N-1$ dimensions can be expressed as a
product of $\zeta_{D}$ and a function $f(\xi_{jk},\zeta_{D}^{2})$,
we can rewrite it as a product of a base function $Q_{0}$ and a
generalized radial function depending upon the first set of
internal variables. Thus, we compare this state with the state of
zero angular momentum in $D=N+1$ dimensions, the number of the
generalized radial function (one), the internal variables (the
first set), and the generalized radial equation (see the proof for
the cases of $D\geq N$) are all the same, respectively. Therefore,
we obtain an exceptional interdimensional degeneracy between these
two states.

In this Letter we have provided a systematic procedure for
analysis of observed degeneracies among different states in
different dimensions and yielded considerable insight into the
energy spectra of an $N$-body system. Since the generalized radial
equations for a quantum $N$-body system in an arbitrary
$D$-dimensional space with a spherically symmetric potential $V$
are derived without any approximation \cite{gu3}, the
interdimensional degeneracies given here are exact
and general.

\vspace{5mm}
\noindent

{\bf ACKNOWLEDGMENTS}. This work was supported by the National
Natural Science Foundation of China.


\begin{thebibliography}{99}

\bibitem{van} J. H. Van Vleck, in Wave Mechanics, the First Fifty Years,
edited by W. C. Price, et al. (Butterworths, London, 1973), pp. 26-37.

\bibitem{her2} D. R. Herrick , J. Math. Phys. {\bf 16}, 281 (1975).

\bibitem{her3} D. R. Herrick and F. H. Stillinger, Phys. Rev.
A {\bf 11}, 42 (1975).

\bibitem{sch} C. Schwartz, Phys. Rev. {\bf 123}, 1700 (1961).

\bibitem{dun2} M. Dunn and D. K. Watson, Ann. Phys. {\bf 251}, 266 (1996).

\bibitem{dun3} M. Dunn and D. K. Watson, Ann. Phys. {\bf 251}, 319 (1996).

\bibitem{dun4} M. Dunn and D. K. Watson, Few-Body Systems {\bf 21}, 187 (1996).

\bibitem{dun5} M. Dunn and D. K. Watson, Phys. Rev. A{\bf 59}, 1109 (1999).

\bibitem{gu1} Xiao-Yan Gu, Bin Duan and Zhong-Qi Ma,
Phys. Rev. A {\bf 64}, 042108 (2001).

\bibitem{gu2} Xiao-Yan Gu, Bin Duan and Zhong-Qi Ma, J. Math.
Phys. {\bf 43}, 2895 (2002).

\bibitem{gu3} Xiao-Yan Gu, Zhong-Qi Ma and Jian-Qiang Sun,
J. Math. Phys. {\bf 44}, 3763 (2003).

\bibitem{gu4} Xiao-Yan Gu, Bin Duan and Zhong-Qi Ma, Phys. Lett.
A {\bf 307}, 55 (2003).


\end{thebibliography}
\end{document}